\documentstyle[aps,prd,multicol]{revtex}
\tighten

\def\Lrule{\vspace*{-0.2in}\noindent\vrule width3.4in height.2pt
  depth.2pt \vrule depth0em height.5em}
\def\Rrule{\vspace{-0.1in}\hfill\vrule depth.5em height0pt \vrule
  width3.4in height.2pt depth.2pt\vspace*{-0.1in}}

\begin{document}

\title{A superspace embedding of the Wess-Zumino model}

\author{J. Barcelos-Neto$^a$}

\address{\mbox{}\\
Instituto de F\'{\i}sica\\
Universidade Federal do Rio de Janeiro\\
RJ 21945-970 - Caixa Postal 68528 - Brazil}
\date{\today}

\maketitle
\begin{abstract}
\hfill{\small\bf Abstract\hspace*{1.7em}}\hfill\smallskip
\par
\noindent
We embed the Wess-Zumino (WZ) model in a wider superspace than the
one described by chiral and anti-chiral superfields.

\end{abstract}

\vfill
%\draft command makes pacs numbers print
\smallskip\mbox{}

\bigskip
\begin{multicols}{2}
{\bf 1.} There is a systematic and interesting formalism for
embedding, developed by Batalin, Fradkin, Fradkina, and Tyutin (BFFT)
\cite{BFFT}, where theories with second-class constraints \cite{Dirac}
are transformed into more general (gauge) theories where all
constraints become first-class. The transformation of constraints from
second to first-class is achieved after extending the phase space by
means of auxiliary variables under the general rule that there is one
pair of canonical variables for each second class constraint. The
method is iterative and can stop in the first step \cite{Banerjee} or
can go on indefinitely \cite{Barc1,Park}. In any case, after all
constraints have been transformed into first-class, it is necessary to
look for the Hamiltonian corresponding to this new theory. The method
also permit us to obtain any involutive quantity that has zero Poisson
brackets with all the constraints. The embedding Hamiltonian can be
obtained in this way, starting from the initial canonical Hamiltonian
and iteratively calculating the corresponding corrections.

\medskip
There is another manner to obtain an embedding Hamiltonian, which
consists in using the BFFT method to obtain involutive coordinates
\cite{Barc1}. The canonical Hamiltonian is then rewritten in terms of
these new coordinates that automatically give it the involutive
condition. Of course, the embedding Hamiltonians obtained from these
two different ways are not necessarily equal. This means that for some
specific theory there may exist more than one possible embeddings. It
is also opportune to say, on the other hand, that there are theories
which cannot be embedded \cite{Barc2}.

\medskip
One of the interesting problem that the BFFT method could be
addressed is the covariant quantization of superparticles and
superstrings, that remained opened for a long time. This problem has
been solved in a embedding procedure but differently of the BFFT
method \cite{Nathan}. In fact, the meaning of embedding in field
theory can be taken as much wider than the cases described by the
BFFT method. The important point is that the embedding theory contains
all the physics of the embedded one. We mention, for example, even the
general procedure of supersymmetrization is an example of embedding.

\medskip
We would like to address the present paper to this point of view of
considering the embedding procedure in a wider way. We concentrate on
the WZ model\cite{WZ} in superfield language \cite{Salam,Wess}.
Conventionally, the WZ model is always developed in terms of
chiral and antichiral superfields, that are examples of irreducible
superfields. We shall consider here a kind of embedding where we
describe the WZ model by using a more general superfield
representation. We shall see
that, contrarily to the bosonic nature of the chiral and antichiral
superfields, the general superfield we have to use is fermionic. We
shall also see that there are two possible terms that can figure in
the Lagrangian and having a relative parameter between them. The
consistency of the obtained theory can be verified by showing it has
the same physics of the WZ model. Finally, we know that a
characteristic of embedding theories is that they have more symmetries
than the embedded one. The same also occurs here. We can show that for
an specific value of the relative parameter between the two terms of
the Lagrangian, there is a kind of gauge symmetry relating all the
fields of the theory.

\bigskip
{\bf 2.} In order to fix the notation and make future comparisons,
let us write down the general form of the real and scalar superfield,

\begin{eqnarray}
\Phi(x,\theta)&=&A(x)+\bar\theta\psi(x)
+\frac{1}{2}\,\bar\theta\theta\,B(x)
\nonumber\\
&&+\frac{i}{2}\,\bar\theta\gamma_5\theta\,C(x)
+\frac{1}{2}\,\bar\theta\gamma^\mu\gamma_5\theta\,A_\mu(x)
\nonumber\\
&&+\frac{1}{2}\,\bar\theta\theta\bar\theta\,\lambda(x)
+\frac{1}{4}\,(\bar\theta\theta)^2\,D(x)\,,
\label{1}
\end{eqnarray}

\noindent
Here, all the spinors are Majorana and are in the Majorana
representation (their components are real). We observe that it
contains eight bosonic and eight fermionic degrees of freedom. We are
going to work in four component notation for the spinor fields. In the
Appendix, we give more details about the notation and convention we
are using and list some useful identities.

\medskip
The irreducible positive and negative chiral superfields,
that contains just four components, are given by

\begin{eqnarray}
\Phi_+(x,\theta)&=&\phi(x)
+\frac{i}{2}\,\bar\theta\gamma^\mu\gamma_5\theta\partial_\mu\phi
-\frac{1}{8}\,(\bar\theta\theta)^2\,\Box\phi
\nonumber\\
&&+\frac{1}{2}\,\bar\theta(1+\gamma_5)\,\psi(x)
-\frac{i}{4}\,\bar\theta\theta\bar\theta
\gamma^\mu(1+\gamma_5)\,\partial_\mu\psi
\nonumber\\
&&+\frac{1}{4}\,\bar\theta(1+\gamma_5)\theta\,F(x)\,,
\label{2}\\
\Phi_-(x,\theta)&=&\phi^\ast(x)
-\frac{i}{2}\,\bar\theta\gamma^\mu\gamma_5\theta\partial_\mu\phi^\ast
-\frac{1}{8}\,(\bar\theta\theta)^2\,\Box\phi^\ast
\nonumber\\
&&+\frac{1}{2}\,\bar\theta(1-\gamma_5)\,\psi(x)
-\frac{i}{4}\,\bar\theta\theta\bar\theta
\gamma^\mu(1-\gamma_5)\,\partial_\mu\psi
\nonumber\\
&&+\frac{1}{4}\,\bar\theta(1-\gamma_5)\theta\,F^\ast(x)\,.
\label{3}
\end{eqnarray}

\noindent
The WZ model \cite{WZ,Wess} is directly obtained (up to some
general constant factor) from an action given by the product of
positive and negative chiral superfields, $S=\int
d^4xd^4\theta\,\Phi_+\Phi_-$.

\bigskip
{\bf 3.} We first observe that the formulation of a supersymmetric
theory, using general superfields and that contains the WZ model as a
particular case, cannot be done in terms of covariant derivatives over
the scalar superfield. This is so because it would violate the correct
mass dimension of the superfield Lagrangian, that should be two. The
correct way is starting from a fermionic superfield, whose general
form reads

\begin{eqnarray}
\Psi_\alpha(x,\theta)&=&\chi_\alpha(x)+\theta_\alpha \phi(x)
+\frac{1}{2}\,\bar\theta\theta\,\psi_\alpha(x)
\nonumber\\
&&+\frac{i}{2}\,\bar\theta\gamma_5\theta\,\lambda_\alpha(x)
+\frac{1}{2}\,\bar\theta\gamma^\mu\gamma_5\theta\,\psi_{\mu\alpha}(x)
\nonumber\\
&&+\frac{1}{2}\,\bar\theta\theta\theta_\alpha F(x)
+\frac{1}{4}\,(\bar\theta\theta)^2\,\eta_\alpha(x)\,.
\label{4}
\end{eqnarray}

\noindent
Consequently, the form of $\bar\Psi_\alpha(x,\theta)$ reads

\begin{eqnarray}
\bar\Psi_\alpha(x,\theta)&=&\bar\chi_\alpha(x)
+\bar\theta_\alpha \phi^\ast(x)
+\frac{1}{2}\,\bar\theta\theta\,\bar\psi_\alpha(x)
\nonumber\\
&&+\frac{i}{2}\,\bar\theta\gamma_5\theta\,\bar\lambda_\alpha(x)
+\frac{1}{2}\,\bar\theta\gamma^\mu\gamma_5\theta\,
\bar\psi_{\mu\alpha}(x)
\nonumber\\
&&+\frac{1}{2}\,\bar\theta\theta\bar\theta_\alpha F^\ast(x)
+\frac{1}{4}\,(\bar\theta\theta)^2\,\bar\eta_\alpha(x)\,.
\label{5}
\end{eqnarray}

\noindent
If we consider the fermionic superfield with mass dimension
$\frac{1}{2}$, the mass dimension of the component fields are

\begin{eqnarray}
&&[\chi]=\frac{1}{2}\,,\hspace{.5cm}
[\phi]=1\,,\hspace{.5cm}
[\psi]=[\lambda]=[\psi_\mu]=\frac{3}{2}\,,
\nonumber\\
&&[F]=2\,,\hspace{.5cm}
[\eta]=\frac{5}{2}\,.
\label{6}
\end{eqnarray}

\noindent
Notice that actually $\phi$, $\psi$, and $F$ have the same mass
dimensions of the corresponding fields of the WZ model.

\medskip
The supersymmetry transformations of the component fields can be
directly obtained by the general supersymmetry transformation relation

\begin{equation}
\delta\Psi_\alpha=(\bar\xi Q)\,\Psi_\alpha\,,
\label{7}
\end{equation}

\noindent
which leads to

\begin{eqnarray}
\delta\chi_\alpha&=&\xi_\alpha \phi\,,
\nonumber\\
\delta \phi&=& -\frac{i}{4}\bar\xi\partial\!\!\!\slash\chi
-\frac{1}{4}\bar\xi\psi
-\frac{i}{4}\bar\xi\gamma_5\lambda
+\frac{1}{4}\bar\xi\gamma_5\gamma^\mu\psi_\mu\,,
\nonumber\\
\delta\psi_\alpha&=&\frac{i}{2}
\bigl(\partial\!\!\!\slash \phi\xi\bigr)_\alpha
+\frac{1}{2}\xi_\alpha F\,,
\nonumber\\
\delta\lambda_\alpha&=&\frac{1}{2}
\bigl(\gamma_5\partial\!\!\!\slash \phi\xi\bigr)_\alpha
+\frac{i}{2}\bigl(\gamma_5\xi\bigr)_\alpha F\,,
\nonumber\\
\delta\psi_{\mu\alpha}&=&\frac{i}{2}
\bigl(\gamma_5\gamma_\mu\partial\!\!\!\slash \phi\xi\bigr)_\alpha
-\frac{1}{2}\bigl(\gamma_5\gamma_\mu\xi\bigr)_\alpha F\,,
\nonumber\\
\delta F&=&\frac{i}{4}\bar\xi\partial\!\!\!\slash\psi
+\frac{1}{4}\bar\xi\gamma_5\partial\!\!\!\slash\lambda
+\frac{i}{4}\bar\xi\gamma^\mu\gamma^\nu\gamma_5\partial_\mu\psi_\nu
-\frac{1}{4}\bar\xi\eta\,,
\nonumber\\
\delta\eta_\alpha&=&\frac{i}{2}
\Bigl(\partial\!\!\!\slash F\xi\Bigr)_\alpha\,.
\label{8}
\end{eqnarray}

\noindent
We observe that the usual transformations of the component fields of
the WZ model are embodied in (\ref{8}).

\medskip
Before going on, it is opportune to make a comment about the number of
bosonic and fermionic degrees of freedom that appear in (\ref{4}) and
(\ref{5}). At first sight, they are not the same. There are thirty two
fermionic degrees of freedom and apparently much less bosonic ones.
What happens is that the bosonic quantities $\phi$ and $F$ are not
representing just single fields. Instead the quantity $\theta_\alpha
\phi$, we must read more generically \cite{Ferrara}

\begin{eqnarray}
\theta_\alpha \phi&\longrightarrow&\theta_\alpha \phi
+(\gamma_5\theta)_\alpha\tilde \phi
+(\gamma^\mu\theta)_\alpha A_\mu
\nonumber\\
&&\phantom{\theta_\alpha \phi}
+(\gamma_5\gamma^\mu\theta)_\alpha\tilde A_\mu
+\frac{1}{2}(\sigma^{\mu\nu}\theta)_\alpha B_{\mu\nu}
\label{9}
\end{eqnarray}

\noindent
that corresponds to sixteen degrees of freedom and the same occurs for
the term with $F$.

\bigskip
{\bf 4.} The most general supersymmetric Lagrangian density expressed
in terms of the spinor superfields, not containing high derivatives
and nonlocal terms, has the form (up to some overall constant factor)

\begin{equation}
{\cal L}=\bar D_\alpha\Psi_\alpha D_\beta\bar\Psi_\beta
+a\,\bar D_\alpha\Psi_\beta D_\alpha\bar\Psi_\beta\,,
\label{10}
\end{equation}

\noindent
where $a$ is a relative normalization parameter that shall be
conveniently fixed. After a long algebraic calculation, we obtain that
the $(\bar\theta\theta)^2$ component of $\cal L$ is given by

\begin{eqnarray}
{\cal L}_{(\bar\theta\theta)^2}
&=&\Bigl(2a+\frac{1}{2}\Bigr)\bar\psi\eta
-\Bigl(2a+\frac{1}{2}\Bigr)\bar\lambda\partial^\mu\psi_\mu
\nonumber\\
&&-\Bigl(a+\frac{1}{4}\Bigr)\bar\psi\Box\chi
+(a-1)FF^\ast-a\phi\Box\phi^\ast
\nonumber\\
&&+\frac{i}{4}\bar\psi\partial\!\!\!\slash\psi
+\frac{1}{2}\bar\psi\gamma_5\partial\!\!\!\slash\lambda
+\frac{i}{2}\bar\psi\gamma_5\partial^\mu\psi_\mu
\nonumber\\
&&+\frac{i}{2}\bar\lambda\gamma_5\eta
+\frac{i}{4}\bar\lambda\partial\!\!\!\slash\lambda
+\frac{1}{2}\bar\psi_\mu\gamma^\mu\gamma_5\eta
\nonumber\\
&&+\frac{i}{4}\bar\psi_\mu\gamma^\mu\gamma^\nu
\partial\!\!\!\slash\psi_\nu
+\frac{i}{2}\bar\eta\partial\!\!\!\slash\chi
-\frac{i}{4}\bar\chi\gamma_5\Box\lambda
\nonumber\\
&&+\frac{1}{4}\bar\chi\gamma_5\gamma^\mu\gamma^\nu
\partial\!\!\!\slash\partial_\mu\psi_\nu\,.
\label{11}
\end{eqnarray}

\noindent
We notice that the bosonic quantities $\phi$ and $F$ do not mix with
any fermionic fields and their equations of motion are the usual ones
that appear in the WZ model (up to the generic scaling parameter
$a$). Concerning the equations of motion for the fermionic fields, we
have

\end{multicols}
\Lrule

\begin{equation}
2(4a+1)\,\eta-(4a+1)\Box\chi+2i\partial\!\!\!\slash\psi
+2\gamma_5\partial\!\!\!\slash\lambda
+2i\gamma_5\partial^\mu\psi_\mu=0\,,
\label{12}
\end{equation}

\begin{equation}
(4a+1)\,\psi+i\gamma_5\lambda-\gamma_5\gamma^\mu\psi_\mu
+i\partial\!\!\!\slash\chi=0\,,
\label{13}
\end{equation}

\begin{equation}
2(4a+1)\,\partial^\mu\psi_\mu+2\gamma_5\partial\!\!\!\slash\psi
-2i\gamma_5\eta-2i\partial\!\!\!\slash\lambda+i\gamma_5\Box\chi=0\,,
\label{14}
\end{equation}

\begin{equation}
2(4a+1)\,\partial^\mu\lambda-2i\gamma_5\partial^\mu\psi
+2\gamma^\mu\gamma_5\eta
+i(\gamma^\mu\gamma^\rho\gamma^\nu
+\gamma^\nu\gamma^\mu\gamma^\rho)\partial_\nu\psi_\rho
+\gamma_5\gamma^\nu\gamma^\mu\partial\!\!\!\slash\partial_\nu\chi=0\,,
\label{15}
\end{equation}

\begin{equation}
(4a+1)\,\Box\psi-2i\partial\!\!\!\slash\eta
+i\gamma_5\Box\lambda
-\gamma_5\gamma^\mu\gamma^\nu
\partial\!\!\!\slash\partial_\mu\psi_\nu=0\,.
\label{16}
\end{equation}

\Rrule
\begin{multicols}{2}

As it has been emphasized, the procedure of embedding must not affect
the physics we already know for the initial theory. We can verify that
this is actually the case by combining these equations to obtain
equations of motion for each component field. For example, by using
(\ref{13}) and (\ref{14}) we eliminate $\lambda$ and $\eta$ from the
remaining equations. The result is

\begin{eqnarray}
&&4a\,\partial\!\!\!\slash\psi-2(2a+1)\gamma_5\partial^\mu\psi_\mu
+\gamma_5\partial\!\!\!\slash\gamma^\mu\psi_\mu+i\Box\chi=0\,,
\label{17}\\
&&2(2a-1)\,\partial\!\!\!\slash\psi+4\gamma_5\partial^\mu\psi_\mu
+\gamma_5\partial\!\!\!\slash\gamma^\mu\psi_\mu+i\Box\chi=0\,,
\label{18}\\
&&\partial\!\!\!\slash\psi+\gamma_5\partial^\mu\psi_\mu=0\,.
\label{19}
\end{eqnarray}

\noindent
The analysis of these equations shows us that for $a=-2$ they are not
independent. On the other hand, for $a\neq-2$ we get

\begin{eqnarray}
&&\partial\!\!\!\slash\psi=0\,,
\label{20}\\
&&\partial^\mu\psi_\mu=0\,,
\label{21}\\
&&\Box\chi-i\gamma_5\partial\!\!\!\slash\gamma^\mu\psi_\mu=0\,.
\label{22}
\end{eqnarray}

\noindent
Introducing these results into (\ref{13}) and (\ref{14}), one
obtains the following equations involving $\lambda$ and $\eta$

\begin{eqnarray}
&&\partial\!\!\!\slash\lambda=0\,,
\label{23}\\
&&\Box\chi-2\eta=0\,.
\label{24}
\end{eqnarray}

We then observe that the equations of motion of the WZ model are
obtained among all the equations of the general model. Hence, the WZ
model is actually embedded in the Lagrangian (\ref{11}). However, this
compatibility with the WZ model did not permit us to completely fix
the relative parameter $a$. It just says it has to be different from
$-2$. We also observe that it cannot be one because it would rule out
the term in $FF^\ast$ of the Lagrangian (\ref{11}).

\bigskip
{\bf 5.} Let us now see that the embedding Lagrangian (\ref{11})
exhibits a kind of gauge symmetry for a specific value of the
parameter $a$. Taking a generic variation of the Lagrangian we obtain

\end{multicols}
\Lrule

\begin{eqnarray}
\delta{\cal L}&=&\delta\bar\psi\,\Bigl[
\Bigl(2a+\frac{1}{2}\Bigr)\,\eta
-\Bigl(a+\frac{1}{4}\Bigr)\,\Box\chi
+\frac{i}{2}\,\partial\!\!\!\slash\psi
+\frac{1}{2}\,\gamma_5\partial\!\!\!\slash\lambda
+\frac{i}{2}\,\gamma_5\partial^\mu\psi_\mu\Bigr]
\nonumber\\
&&+\delta\bar\eta\Bigl[
\Bigl(2a+\frac{1}{2}\Bigr)\,\psi
+\frac{i}{2}\,\gamma_5\lambda
-\frac{1}{2}\,\gamma_5\gamma^\mu\psi_\mu
+\frac{i}{2}\,\partial\!\!\!\slash\chi\Bigr]
\nonumber\\
&&-\delta\bar\lambda\Bigl[
\Bigl(2a+\frac{1}{2}\Bigr)\,\partial^\mu\psi_\mu
+\frac{1}{2}\gamma_5\partial\!\!\!\slash\psi
-\frac{i}{2}\gamma_5\eta
-\frac{i}{2}\partial\!\!\!\slash\lambda
+\frac{i}{4}\gamma_5\Box\chi\Bigr]
\nonumber\\
&&+\delta\bar\psi_\mu\Bigl[
\Bigl(2a+\frac{1}{2}\Bigr)\,\partial^\mu\lambda
-\frac{i}{2}\,\gamma_5\partial^\mu\psi
+\frac{1}{2}\,\gamma^\mu\gamma_5\eta
+\frac{i}{4}\,(\gamma^\mu\gamma^\rho\gamma^\nu
+\gamma^\nu\gamma^\mu\gamma^\rho)\partial_\nu\psi_\rho
+\frac{1}{4}\,\gamma_5\gamma^\nu\gamma^\mu
\partial\!\!\!\slash\partial_\nu\chi\Bigr]
\nonumber\\
&&-\delta\bar\chi\Bigl[
\Bigl(a+\frac{1}{4}\Bigr)\,\Box\psi
-\frac{i}{2}\,\partial\!\!\!\slash\eta
+\frac{i}{4}\,\gamma_5\Box\lambda
-\frac{1}{4}\,\gamma_5\gamma^\mu\gamma^\nu
\partial\!\!\!\slash\partial_\mu\psi_\nu\Bigr]
+\delta A^\ast\Box A
+\delta F^\ast F\,.
\label{25}
\end{eqnarray}

\Rrule
\begin{multicols}{2}
\noindent
Looking at the equation of motion (\ref{22}), it suggests us that
a possible gauge transformation for $\bar\psi_\mu$ should have the
form

\begin{equation}
\delta\bar\psi_\mu(x)=\bar\alpha(x)\gamma_\mu\gamma_5\,,
\label{26}
\end{equation}

\noindent
where $\alpha(x)$ is a Majorana spinor that plays the role of a gauge
parameter. Keeping in mind the mass dimension of the fields that
appear in (\ref{25}), we infer that the gauge transformations for
the remaining fields should be

\begin{eqnarray}
\delta\bar\psi(x)&=&b\,\bar\alpha(x)\,,
\nonumber\\
\delta\bar\eta(x)&=&c\,\partial_\mu\bar\alpha(x)\gamma^\mu\,,
\nonumber\\
\delta\bar\lambda(x)&=&d\,\bar\alpha(x)\gamma_5\,,
\nonumber\\
\delta\bar\chi(x)&=&e\,\frac{1}{\Box}\partial_\mu\bar\alpha(x)
\gamma^\mu\,,
\nonumber\\
\delta A^\ast(x)&=&0\,,
\nonumber\\
\delta F^\ast(x)&=&0\,,
\label{27}
\end{eqnarray}

\noindent
where $b$, $c$, $d$, and $e$ are parameters to be conveniently fixed.
Replacing (\ref{26}) and (\ref{27}) into (\ref{25}) we get that
the necessary condition to get the symmetry is

\begin{eqnarray}
&&2b+2ic+2id-ie=8a+2\,,
\nonumber\\
&&2b+(4a+1)2ic+2id-(4a+1)ie=2\,,
\nonumber\\
&&(4a+1)b+id-ie=4\,,
\nonumber\\
&&(4a+1)b+2ic+id=-2\,.
\label{28}
\end{eqnarray}

\noindent
These correspond to the coefficients of
$\bar\alpha\gamma_5\partial\!\!\!\slash\lambda$,
$\bar\alpha\partial\!\!\!\slash\psi$, $\bar\alpha\eta$, and
$\bar\alpha\Box\chi$, respectively. There is still another equation
to be verified which is related to the field $\psi_\mu$, namely

\begin{eqnarray}
&&(-6i+2c+e)\,\bar\alpha\gamma_5\gamma^\mu\partial\!\!\!\slash\psi_\mu
\nonumber\\
&&\phantom{(-6i}
+[4i+2bi-4c-2d(4a+1)]\,\bar\alpha\gamma_5\partial^\mu\psi_\mu=0\,,
\label{29}
\end{eqnarray}

\noindent
where one cannot infer any conclusion for the coefficients of
$\bar\alpha\gamma_5\gamma^\mu\partial\slash\!\!\!\psi_\mu$ and
$\bar\alpha\gamma_5\partial^\mu\psi_\mu$ because these terms are not
independent.

\medskip
Considering the set given by (\ref{28}), one can solve it to express
$b$, $c$, $d$, and $e$ in terms of $a$. The result is $b=-1$, $c=2i$,
$d=-i(4a+3)$, and $e=2i$ (it is important to mention that this
solution exists only if $a\neq0$). Introducing now this result into
(\ref{29}), we get a providential cancelation of the first term. The
second one becomes

\begin{equation}
a(a+1)\,\bar\alpha\gamma_5\partial^\mu\psi_\mu=0\,.
\label{30}
\end{equation}

\noindent
Since $a$ cannot be zero, we get that the symmetry given by
(\ref{26}) and (\ref{27}) fixes the parameter $a$ into $-1$ (this
value is compatible with all the previous boundary conditions).

\bigskip
{\bf 6.} In this work we have embedded the WZ model in a
wider superspace than the one described by chiral and antichiral
superfields. We have show that just the fermionic general superfield
is appropriated to be used and the consistency condition of the
embedding is verified by showing that the same equations of motion of
the WZ model are among the equations of motion of the general model.
Finally, we have also shown that the embedding theory has a kind of
gauge symmetry. This symmetry permit us to fix a relative parameter
that appear in the two terms of the Lagrangian.

\vspace{1cm}
\noindent {\bf Acknowledgment:} This work is supported in part by
Conselho Nacional de Desenvolvimento Cient\'{\i}fico e Tecnol\'ogico
- CNPq with the support of PRONEX 66.2002/1998-9.

\appendix
\renewcommand{\theequation}{A.\arabic{equation}}
\setcounter{equation}{0}
\section{}
%\section{Convention, notation, and identities}

In this Appendix, we present the notation, convention and the main
identities used throughout the paper. The gamma matrices satisfy the
usual relations $\{\gamma^\mu,\gamma^\nu\}=
2\eta^{\mu\nu}$ and $\gamma^\mu=\gamma^0\gamma^{\mu\dagger}\gamma^0$.
We adopt the metric convention
$\eta^{\mu\nu}={\rm diag.}(1,-1,-1,-1)$. We take the completely
antisymmetric tensor $\epsilon^{\mu\nu\rho\lambda}$ given by
$\epsilon^{0123}=1$. The matrices $\gamma_5$ and $\sigma^{\mu\nu}$
are defined as

\begin{eqnarray}
\gamma_5&=&i\gamma^0\gamma^1\gamma^2\gamma^3\,,
\label{A.1}\\
\sigma^{\mu\nu}&=&\frac{i}{2}[\gamma^\mu,\gamma^\nu]\,.
\label{A.2}
\end{eqnarray}

Let us list below some useful identities involving gamma matrices

\begin{eqnarray}
\gamma^\mu\gamma^\nu\gamma^\rho&=&\eta^{\mu\nu}\gamma^\rho
-\eta^{\mu\rho}\gamma^\mu+\eta^{\nu\rho}\gamma^\mu
-i\epsilon^{\mu\nu\rho\lambda}\gamma_5\gamma_\lambda\,,
\label{A.3}\\
\gamma_5\gamma^\mu\gamma^\nu&=&\eta^{\mu\nu}\gamma_5
+\frac{1}{2}\,\epsilon^{\mu\nu\rho\lambda}\sigma_{\rho\lambda}\,,
\label{A.4}\\
\gamma_5\sigma^{\mu\nu}&=&\frac{i}{2}\epsilon^{\mu\nu\rho\lambda}
\sigma_{\rho\lambda}\,,
\label{A.5}\\
\gamma^\mu\sigma^{\rho\lambda}
&=&\frac{i}{2}(\eta^{\mu\rho}\gamma^\lambda
-\eta^{\mu\lambda}\gamma^\rho)
+\frac{1}{2}\epsilon^{\mu\rho\lambda\nu}\gamma_5\gamma_\nu\,,
\label{A.6}\\
\sigma^{\mu\nu}\gamma^\rho
&=&\frac{i}{2}(\eta^{\nu\rho}\gamma^\mu
-\eta^{\mu\rho}\gamma^\nu)
+\frac{1}{2}\epsilon^{\mu\nu\rho\lambda}\gamma_5\gamma_\lambda\,,
\label{A.7}\\
\sigma^{\mu\nu}\sigma^{\rho\lambda}
&=&\frac{i}{4}\epsilon^{\mu\nu\rho\lambda}\gamma_5
+\frac{1}{4}(\eta^{\mu\rho}\eta^{\nu\lambda}
-\eta^{\mu\lambda}\eta^{\nu\rho})
\nonumber\\
&&-\frac{i}{2}(\eta^{\mu\rho}\eta^{\nu\alpha}\eta^{\lambda\beta}
+\eta^{\nu\lambda}\eta^{\mu\alpha}\eta^{\rho\beta}
-\rho\leftrightarrow\lambda)\,.
\label{A.8}
\end{eqnarray}

\noindent
Further

\begin{eqnarray}
{\rm tr}\,\gamma^\mu\gamma^\nu&=&4\eta^{\mu\nu}\,,
\nonumber\\
{\rm tr}\,\gamma_5&=&0\,,
\nonumber\\
{\rm tr}\,\gamma_5\gamma^\mu\gamma^\nu&=&0\,,
\nonumber\\
{\rm tr}\,\gamma_5\gamma^\mu\gamma^\nu\gamma^\rho\gamma^\lambda
&=&4i\epsilon^{\mu\nu\rho\lambda}\,,
\nonumber\\
{\rm tr}\,\sigma^{\mu\nu}\sigma^{\rho\lambda}
&=&4(\eta^{\mu\rho}\eta^{\nu\lambda}-\eta^{\mu\lambda}\eta^{\nu\rho})
\,.
\label{A.9}
\end{eqnarray}

\noindent
Considering $\psi$ and $\chi$ as Majorana spinors, we also have

\begin{eqnarray}
\bar\psi\chi&=&\bar\chi\psi\,,
\label{A.10}\\
\bar\psi\gamma_5\chi&=&\bar\chi\gamma_5\psi\,,
\label{A.11}\\
\bar\psi\gamma^\mu\gamma_5\chi&=&\bar\chi\gamma^\mu\gamma_5\psi\,,
\label{A.12}\\
\bar\psi\gamma^\mu\chi&=&-\bar\chi\gamma^\mu\psi\,,
\label{A.13}\\
\bar\psi\sigma^{\mu\nu}\chi&=&-\bar\chi\sigma^{\mu\nu}\psi\,,
\label{A.14}\\
\bar\psi\gamma^\mu\gamma^\nu\chi&=&\bar\chi\gamma^\nu\gamma^\mu\psi\,.
\label{A.15}
\end{eqnarray}

\noindent
Using the relations (\ref{A.3})-(\ref{A.15}), we obtain additional
relations

\begin{eqnarray}
\bar\psi\gamma_5\sigma^{\mu\nu}\chi
&=&-\bar\chi\gamma_5\sigma^{\mu\nu}\psi\,,
\nonumber\\
\bar\psi\gamma_5\gamma^\mu\gamma^\nu\chi
&=&\bar\chi\gamma_5\gamma^\nu\gamma^\mu\psi\,,
\nonumber\\
\bar\psi\gamma^\mu\gamma^\nu\gamma^\rho\chi
&=&-\bar\chi\gamma^\rho\gamma^\nu\gamma^\mu\psi\,,
\nonumber\\
\bar\psi\gamma_5\gamma^\mu\gamma^\nu\gamma^\rho\chi
&=&\bar\chi\gamma_5\gamma^\rho\gamma^\nu\gamma^\mu\psi\,,
\nonumber\\
\bar\psi\gamma_5\gamma^\rho\sigma^{\mu\nu}\chi
&=&-\bar\chi\gamma_5\sigma^{\mu\nu}\gamma^\rho\psi\,,
\nonumber\\
\bar\psi\sigma^{\mu\nu}\gamma^\rho\gamma^\lambda\chi
&=&\bar\chi\gamma^\lambda\gamma^\rho\sigma^{\mu\nu}\psi\,,
\nonumber\\
\bar\psi\gamma_5\sigma^{\mu\nu}\gamma^\alpha\sigma^{\rho\lambda}\chi
&=&\bar\chi\gamma_5\sigma^{\rho\lambda}\gamma^\alpha\sigma^{\mu\nu}
\psi\,.
\label{A.16}
\end{eqnarray}

The Fierz identity reads

\begin{equation}
\frac{1}{4}(\Gamma^A)_{\alpha\beta}(\Gamma_A)_{\sigma\rho}
=\delta_{\alpha\rho}\delta_{\beta\sigma}\,,
\label{A.17}
\end{equation}

\noindent
where $\Gamma^A$ is generically representing the independent
matrices: $\Gamma^1=1$, $\Gamma^2$ to $\Gamma^5=\gamma^\mu$,
$\Gamma^6=\gamma_5$, $\Gamma^7$ to $\Gamma^{10}=\gamma^\mu\gamma_5$,
$\Gamma^{11}$ to $\Gamma^{16}=\sigma^{\mu\nu}$. Concerning to
$\Gamma_A$, the corresponding relations are almost trivial,
we just have to notice the inverse order between $\gamma_5$ and
$\gamma_\mu$ from $\Gamma_7$ to $\Gamma_{10}=\gamma_5\gamma_\mu$.
Using the Fierz identity, we obtain

\begin{eqnarray}
\theta_\alpha\bar\theta_\beta
&=&-\frac{1}{4}\delta_{\alpha\beta}\bar\theta\theta
-\frac{1}{4}\gamma_{5\alpha\beta}\bar\theta\gamma_5\theta
\nonumber\\
&&-\frac{1}{4}(\gamma^\mu\gamma_5)_{\alpha\beta}
\bar\theta\gamma_5\gamma_\mu\theta\,,
\nonumber\\
\bar\theta\gamma_5\theta\,\bar\theta_\alpha
&=&-\bar\theta\theta\,(\bar\theta\gamma_5)_\alpha\,,
\nonumber\\
\theta_\alpha\,\bar\theta\gamma_5\theta
&=&-(\gamma_5\theta)_\alpha\bar\theta\theta\,,
\nonumber\\
\bar\theta\gamma_5\gamma_\mu\theta\,\bar\theta_\alpha
&=&-\bar\theta\theta\,(\bar\theta\gamma_5\gamma_\mu)_\alpha\,,
\nonumber\\
\theta_\alpha\,\bar\theta\gamma_5\gamma_\mu\theta
&=&-(\gamma_5\gamma_\mu\theta)_\alpha\bar\theta\theta\,,
\nonumber\\
\bar\theta\gamma_5\theta\,\bar\theta\gamma_5\theta
&=&-\,(\bar\theta\theta)^2\,,
\nonumber\\
\bar\theta\gamma^\mu\gamma_5\theta\,
\bar\theta\gamma^\nu\gamma_5\theta
&=&\eta^{\mu\nu}(\bar\theta\theta)^2\,,
\nonumber\\
\bar\theta\theta\,\bar\theta\gamma_5\theta&=&0\,,
\nonumber\\
\bar\theta\theta\,\bar\theta\gamma^\mu\gamma_5\theta&=&0\,,
\nonumber\\
\bar\theta\gamma_5\theta\,\bar\theta\gamma^\mu\gamma_5\theta&=&0\,.
\label{A.18}
\end{eqnarray}

The supersymmetry charge and derivative operators are defined by

\begin{eqnarray}
Q_\alpha&=&\frac{\partial}{\partial\bar\theta_\alpha}
+i(\gamma^\mu\theta)_\alpha\partial_\mu\,,
\nonumber\\
\bar Q_\alpha&=&-\frac{\partial}{\partial\theta_\alpha}
-i(\bar\theta\gamma^\mu)_\alpha\partial_\mu\,,
\nonumber\\
D_\alpha&=&\frac{\partial}{\partial\bar\theta_\alpha}
-i(\gamma^\mu\theta)_\alpha\partial_\mu\,,
\nonumber\\
\bar D_\alpha&=&-\frac{\partial}{\partial\theta_\alpha}
+i(\bar\theta\gamma^\mu)_\alpha\partial_\mu\,.
\label{A.19}
\end{eqnarray}

Positive and negative chiralities are defined as

\begin{equation}
\theta_{\pm}=\frac{1}{2}(1\pm\gamma_5)\theta\,,
\label{A.20}
\end{equation}

\noindent
consequently,

\begin{eqnarray}
\frac{\partial}{\partial\theta_\pm}\theta_\pm
&=&\frac{1}{2}(1\pm\gamma_5)\,,
\nonumber\\
\frac{\partial}{\partial\bar\theta_\pm}\bar\theta_\pm
&=&\frac{1}{2}(1\mp\gamma_5)\,,
\nonumber\\
\frac{\partial}{\partial\theta_\pm}\bar\theta_\pm
&=&\frac{1}{2}(1\pm\gamma_5)\gamma^0\,,
\nonumber\\
\frac{\partial}{\partial\bar\theta_\pm}\theta_\pm
&=&\frac{1}{2}(1\mp\gamma_5)\gamma^0\,,
\nonumber\\
\frac{\partial}{\partial\theta_\pm}\theta_\mp&=&0\,,
\nonumber\\
\frac{\partial}{\partial\bar\theta_\pm}\bar\theta_\mp&=&0\,,
\nonumber\\
\frac{\partial}{\partial\theta_\pm}\bar\theta_\mp&=&0\,,
\nonumber\\
\frac{\partial}{\partial\bar\theta_\pm}\theta_\mp&=&0\,.
\label{A.21}
\end{eqnarray}

\noindent
The positive and negative chiral superfields satisfy

\begin{equation}
D_\mp\Phi_\pm=0\,.
\label{A.22}
\end{equation}

\end{multicols}

\vspace{1cm}
\begin{thebibliography}{30}
\bibitem[a]{}e-mail: {\tt barcelos@ if.ufrj.br}
\vspace{.5cm}
\bibitem{BFFT}I.A. Batalin and E.S. Fradkin, Phys. Lett. B180 (1986)
157; Nucl. Phys.  B279 (1987) 514; I.A. Batalin, E.S. Fradkin, and
T.E. Fradkina, {\it ibid.} B314 (1989) 158; B323 (1989) 734; I.A.
Batalin and I.V. Tyutin, Int. J. Mod. Phys. A6 (1991) 3255.
\bibitem{Dirac} P.A.M.  Dirac, Can. J. Math. 2 (1950) 129; {\it
Lectures on quantum mechanics} (Yeshiva University, New York, 1964).
\bibitem{Banerjee} See for example, N. Banerjee, R. Banerjee, and S.
Ghosh, Nucl. Phys. B417 (1994) 257 and references therein.
\bibitem{Barc1} R. Banerjee and J. Barcelos-Neto, Nucl. Phys. B499
(1997) 453.
\bibitem{Park} M.-I. Park and Y.-J. Park, Int. J. Mod. Phys. A13
(1998) 2179.
\bibitem{Barc2} J. Barcelos-Neto, Phys. Rev. {\bf D55} (1997) 2265.
\bibitem{Nathan} See for example the recent paper N. Berkovits, JHEP
{\bf 04} (2000) 018 and references therein.
\bibitem{WZ} J. Wess and B. Zumino, Nucl. Phys. {\bf B70} (1974) 39.
\bibitem{Salam} A. Salam and J. Strathdee, Nucl. Phys. {\bf B76}
(1974) 477; Phys. Rev. D11 (1975) 1521.
\bibitem{Wess} J. Wess and J. Bagger, {\it Supersymmetry and
Supergravity}, Princeton Univ. Press (1992) and references therein.
\bibitem{Ferrara} See, for example, S. Ferrara, N. Cim. Lett. {\bf 13}
(1975) 629.

\end{thebibliography}
\end{document}